\documentclass[12pt]{article}
\usepackage{amssymb}
\usepackage{epsfig}
\newcommand{\beq}{\begin{equation}}
\newcommand{\eeq}{\end{equation}}
\newcommand{\beqn}{\begin{eqnarray}}
\newcommand{\eeqn}{\end{eqnarray}}
\newcommand{\bea}[1]{\beq\begin{array}{#1}}
\newcommand{\eea}{\end{array}\eeq}
\newcommand{\eq}[1]{(\ref{#1})}

\newcommand{\tr}{\mathop{\rm Tr}}

\newcommand{\cZ}{{\cal Z}}

\newcommand{\NP}[3]{{\em Nucl. Phys. }{\bf #1} (#2) #3}
\newcommand{\NPPS}[3]{{\em Nucl. Phys. Proc. Suppl. }{\bf #1} (#2) #3}
\newcommand{\PL}[3]{{\em Phys. Lett. }{\bf #1} (#2) #3}

\newcommand{\PR}[3]{{\em Phys. Rev. }{\bf #1} (#2) #3}

\newcommand{\PTPS}[3]{{\em Prog. Theor. Phys. Suppl. }{\bf #1} (#2) #3}

\begin{document}
\date{}
\title{Anatomy of Isolated Monopole in Abelian Projection\\
of $SU(2)$ Lattice Gauge Theory
\vskip-40mm
\rightline{\small KANAZAWA 01-06}
\vskip 40mm
}
\author{V.A. Belavin, M.I. Polikarpov, A.I. Veselov
\\
{\small\it Institute of Theoretical and  Experimental Physics,
B.Cheremushkinskaya 25, Moscow,}\\
{\small\it  117259, Russia}}
\sloppy
\maketitle

\begin{abstract}\noindent
We study the structure of the isolated
static monopoles in the maximal Abelian projection of $SU(2)$ lattice
gluodynamics. Our estimation of the monopole radius is: $R^{mon} \approx
0.06 \,\, fm$.
\end{abstract}

{\bf 1.}
The monopole confinement mechanism in $SU(2)$ lattice gauge theory is
confirmed by many numerical calculations (see {\em e.g.} reviews 
\cite{reviews}). In the maximal Abelian projection monopole currents form 
one big cluster and several small clusters. The big cluster, {\it infrared 
(IR) cluster}, percolates and has a nontrivial fractal dimension, $D_f 
>1$~\cite{percolation}. The properties of small, {\it ultraviolet (UV)}, 
clusters differs much from those of the IR cluster, it can be shown that the 
IR monopole cluster is responsible for the confinement of 
quarks~\cite{clusters}. As it was shown in the recent publication 
\cite{monanat} the structure of IR and UV monopoles is completely different, 
and monopoles in IR clusters are condensed due to their special anatomy. In 
this publication we study the structure of Abelian monopoles in $SU(2)$ 
lattice gauge theory in a different way than it was done in ref. 
\cite{monanat}. We study the structure of the isolated monopoles, the 
results show that nontrivial monopole anatomy plays crucial role in the 
confinement phenomenon.

{\bf 2.} The plaquette action of the
compact electrodynamics (cQED),
\beq\label{ScQED}
S^P_{cQED} = \beta_{U(1)} \cos \theta_P \, ,
\eeq
is close to the action of $SU(2)$ lattice gauge theory in the maximal
Abelian projection at small values of the bare charge $g$ (in the continuum
limit of gluodynamics). The proof is as follows. By definition the
maximal Abelian projection corresponds to the maximization of the functional
$R$ with respect to all gauge transformations $\Omega$:

\beq \label{MAAPdef}
\max_\Omega R[{U}^\Omega_l],\,\,U_l^\Omega = \Omega^+ U_l \Omega,\,\,
R[U_l] = \sum_l Tr[\sigma_3 U_l^+ \sigma_3 U_l] =
\sum_l \cos 2 \varphi_l\,.
\eeq
here we use the standard parametrization of the link matrix, $U_{l,11} =
U_{l,22}^* = \cos \varphi_l e^{i\theta_l}\,,\,\,\, U_{l,12} = U_{l,21}^* =
\sin \varphi_l e^{- i \chi_l}$.  Thus, the maximization of $R$,
eq.~\eq{MAAPdef}, corresponds to the maximization of the modules of the
diagonal elements $U_{l,11}, \, U_{l,22}$.  The $SU(2)$ plaquette action is
$S^P_{SU(2)} = \beta\, \frac 12\tr U_P = \beta \cos \theta_l\cos\varphi_l$,
and at large values of $\beta$ in the maximal Abelian projection
$\cos\varphi_l$ is close to unity (due to \eq{MAAPdef}), $\varphi_l$ is
small and $SU(2)$ plaquette action has the form:

\beq \label{SSU2}
S^P_{SU(2)} = \beta\left[ \cos\theta_P\cos \varphi_1 \cos
\varphi_2 \cos\varphi_3 \cos\varphi_4 + O(\sin\varphi_l)\right].
\eeq

{\bf 3.} The larger value of $\beta$, the smaller $\sin\varphi_l$, and
$S^P_{SU(2)}$ \eq{SSU2} coincides with $S^P_{cQED}$ \eq{ScQED} in the limit
$\beta \to \infty$. On the other hand at small values of the bare charge (at
large values of $\beta_{U(1)}$) the compact electrodynamics is in the
deconfinement phase, and gluodynamics is in the confinement phase; on the
other hand the actions of both theories are close to each other. The
explanation of this paradox was given in Refs.~\cite{compensation,monanat},
it was shown that the action of the non-diagonal gluons, $S^{off}$, on the
plaquettes near the monopole from IR clusters is negative, and the full
non-Abelian action, $S^{SU(2)}=S^{off}+S^{Abel}$, is smaller than the
Abelian part of the action. The standard qualitative proof of the existence
of the deconfinement phase transition in cQED is the representation of the
partition function as the sum over the monopole trajectories of length $L$:

\beq\label{ZQED}
\cZ = \sum_L exp\{ - \beta L c\} ( 7)^L\, ,
\eeq
here $c$ is the action of the unit length of the monopole trajectory, $7^L$ 
is the entropy of the line of the length $L$ drawn on $4D$ hypercubic
lattice. It is clear that at $\beta = \beta_c\equiv \ln 7/c$ there exists
the phase transition in the sum \eq{ZQED}. This phase transition is absent
in lattice gluodynamics since in this case the monopoles have nontrivial
structure and the action of the unit of monopole trajectory in lattice units
depends on $\beta$: $c= c(\beta)$ the sum \eq{ZQED} is always divergent, the
monopoles are condensed and form the percolating cluster. The monopole
condensation was proven in gluodynamics by several independent
calculations~\cite{OrderP}.

{\bf 4.} In ref. \cite{monanat} the average nonabelian action on the
plaquettes near the monopole trajectory in IR clusters has been measured.  
Since the lattice spacing $a$ depends on $\beta$ the calculations at
various $\beta$ correspond to the measurement of the field strength at
various distances, $a(\beta)/2$, from the monopole center. Below we present
the results of another measurement, we calculate the average field strength 
on the plaquettes closest to the monopole center for monopoles which satisfy 
the following two conditions:

{\em (i)} the link with the monopole current has the same direction
as the previous and subsequent monopole current links;

{\em (ii)} there are no other monopoles at the distance less than $2a$ from 
the considered monopole, except of monopoles discussed at point {\em (i)}.

Thus we study ``static'' and ``standing along monopoles'', we call such
monopoles as {\em isolated} monopoles. The results of the calculations are
shown on Figure~1 where we plot the dependence of
$S_{SU(2)}^i=6\beta\cdot\frac12(<\mbox{Tr}U_P^{imon}> - <\mbox{Tr}U_P >)$,
on $a/2$; $U_P^{imon}$ are the plaquette matrices corresponding to plaquettes
closest to the isolated monopole, the normalization of $S_{SU(2)}^i$ is such
that it exactly corresponds to the action of the unit length of the monopole
trajectory. If $S_{SU(2)}^i < \ln 7$ the isolated monopoles are condensed
(see the discussion of the partition function \eq{ZQED}). On Figure~1 we
also show the quantity $S_{SU(2)}^{all} =
6\beta\cdot\frac12(<\mbox{Tr}U_P^{allmon}> - <\mbox{Tr}U_P >)$, here
$U_P^{allmon}$ is the matrix corresponding to plaquettes closest to all 
monopoles (isolated and not isolated).

\begin{figure}
\centerline{\epsfig{file=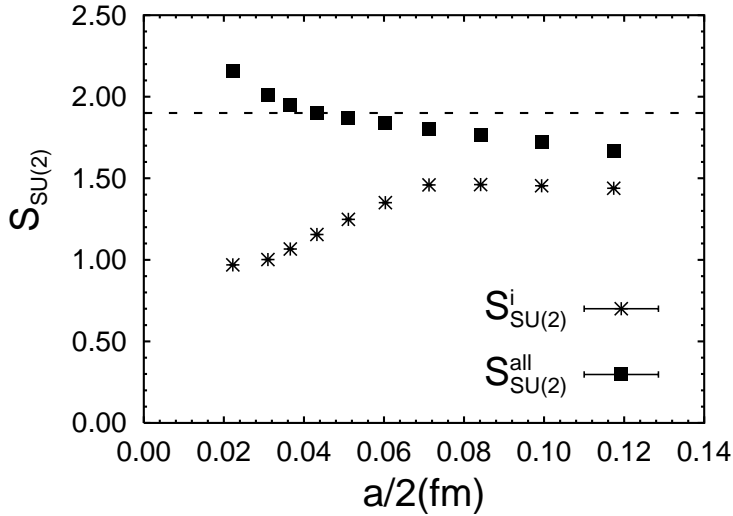,angle=270,width=10cm}}
\caption{The dependence of $S^i_{SU(2)}$ (stars) and $S_{SU(2)}^{all}$
(squares) on $a/2$. The dashed
line corresponds to $S = \ln 7$} \label{nonabeact}.
\end{figure}

The main conclusion from Figure~1 is that the action of
isolated monopoles decreases when we approach the monopole center, and that
these monopoles are condensed. Our numerical results also show that the
{\em abelian} part of the action of the isolated monopoles increases when we
approach the monopole center. Thus the contribution of nondiagonal gluons to
nonabelian action of monopole is negative, and just due to that these
monopoles differs from monopoles in cQED and are condensed at any value of 
$\beta$.

{\bf 5.} Following ref. \cite{monanat} we can estimate the radius of the
isolated monopole, $R_m$, as the point of the maximal derivative of the
function $S_{SU(2)}^i (a/2)$, we thus get: $R_m\approx 0.065\, fm$. Note
that other dimensional numbers which characterize the gluodynamic vacuum are
an order of magnitude larger. For example, the average intermonopole
distance \cite{monanat}, which can be estimated from the results of
ref.~\cite{clusters} is: $R_m\approx 0.5 \,fm$; the width of the abelian
confining flux tube is: $R_t \approx 0.3 \, fm$ \cite{Bali}; the average
instanton radius is:  $R_I \approx 0.3 \, fm$ (see \cite{RI} and references 
therein). Thus we see that in the QCD vacuum there exists a rather small 
scale, $0.065 \, fm$, such small scale was already discussed in various 
studies of QCD vacuum \cite{smallsc,monanat}.

{\bf 6.} M.I.P. and A.I.V. acknowledge the kind hospitality of the staff of
the Institute for Theoretical Physics of Kanazawa University, where the work
was initiated. The work was partially supported by grants,
RFBR 01-02-17456, RFBR 00-15-96786, INTAS 00-00111, JSPS Grant in Aid for
Scientific Research (B) (Grant No. 10440073 and No. 11695029) and CRDF award
RP1-2103.

\end{document}